\documentclass{WileyMSP-template}

\begin{document}

\pagestyle{fancy}
\rhead{\includegraphics[width=2.5cm]{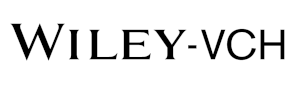}}

\title{Realization of $Z_2$ Topological Metal
in Single-Crystalline Nickel Deficient NiV$_2$Se$_4$}

\maketitle


\author{Sitaram Ramakrishnan*}
\author{Shidaling Matteppanavar}
\author{Andreas Sch\"{o}nleber}
\author{Bikash Patra}
\author{Birender Singh}
\author{Arumugam Thamizhavel}
\author{Bahadur Singh*}
\author{Srinivasan Ramakrishnan*}
\author{Sander van Smaalen*}

\begin{affiliations}
S. Ramakrishnan \\
Department of Quantum Matter, AdSE,
Hiroshima University,
Higashi-Hiroshima 739-8530, Japan\\
Email Address: niranj002@gmail.com

S. Matteppanavar,
B. Patra,
B. Singh,
B. Singh,
A. Thamizhavel,
S. Ramakrishnan \\
Department of Condensed Matter Physics
and Materials Science,
Tata Institute of Fundamental Research,
Mumbai 400005, India \\
Email Address:  bahadur.singh@tifr.res.in, ramky07@gmail.com

A. Sch\"{o}nleber,
S. van Smaalen \\
Laboratory of Crystallography,
University of Bayreuth, 95447 Bayreuth, Germany \\
Email Address:  smash@uni-bayreuth.de

\end{affiliations}



\begin{abstract}
Temperature-dependent electronic and
magnetic properties are reported for
a $Z_2$ topological metal
single-crystalline nickel-deficient NiV$_2$Se$_4$.
It is found to crystallize in the monoclinic
Cr$_3$S$_4$ structure type with space group $I2/m$.
From single-crystal x-ray diffraction, we find that
there are vacancies on the Ni site, resulting in
the composition Ni$_{0.85}$V$_{2}$Se$_{4}$
in agreement with our electron-probe microanalysis.
The electrical resistivity shows metallic behavior
with a broad anomaly around 150--200 K that is also
observed in the heat capacity data.
This anomaly indicates a change of state of the
material below 150 K.
We believe that this anomaly could be
due to spin fluctuations or charge-density-wave 
(CDW) fluctuations,
where the lack of long-range order is caused by
vacancies at the Ni site of Ni$_{0.85}$V$_2$Se$_4$.
Although we fail to observe any structural
distortion in this crystal down to 1.5 K,
its electronic and thermal properties are anomalous.
The observation of non-linear temperature
dependence of resistivity as well as an enhanced
value of the Sommerfeld coefficient
$\gamma = 104.0\,(1)$ mJ/mol{$\cdot$}K$^2$
suggests strong electron-electron correlations in this material.
The first-principles calculations performed
for NiV$_2$Se$_4$, which are also applicable to
Ni$_{0.85}$V$_{2}$Se$_{4}$, classify this
material as a topological metal with
$Z_2 = (1;110)$ and coexisting electron and hole
pockets at the Fermi level.
The phonon spectrum lacks any soft phonon mode,
consistent with the absence of periodic lattice
distortion in the present experiments.
\end{abstract}


\section{Introduction}

Topological metals such as Dirac and Weyl
semimetals represent exotic quantum materials
with non-trivial band crossings at the charge
neutrality point in the bulk band structure
that lead to peculiar quasi-particle
excitations and physical properties.
There has been a surge of interest in
realizing these topological metals, owing to
their novelty for fundamental science and
device applications~\cite{Review2020,singhb2022a}.
Here, we show that nickel-deficient
Ni$_{0.85}$V$_{2}$Se$_{4}$ is a topological
metal with $Z_2 = (1;110)$ based on
first-principles calculations.
Our bulk measurements on single-crystalline
Ni$_{0.85}$V$_{2}$Se$_{4}$ show unusual
quasi-particle excitations.
The presence of vacancies is a common feature
of compounds AT$_2$X$_4$ (A and T are
transition metals; X is chalcogenide),
which display exotic electronic and
magnetic properties.

Cubic thiospinels of the type AT$_2$S$_4$
(A = Cu, Ni; T = V, Rh, Ir) have
attracted attention due to the unusual
ground states exhibited by them
\cite{haginot1995a,nagatas1994a,furubayashit1994a,sekinet1984a,mahyj1987a,radaellipg2002a}.
For instance, superconductivity (e.g.
in CuRh$_2$S$_4$) \cite{haginot1995a},
magnetic order in insulators
(CuIr$_2$S$_4$) \cite{nagatas1994a, furubayashit1994a}
and charge density waves (CDWs; CuV$_2$S$_4$)
\cite{sekinet1984a, mahyj1987a}
have been found to exist in this class
of compounds.
It is worthwhile to recall here that the
CDW in CuV$_2$S$_4$ was suppressed in
as-grown crystals, because of the
presence of defects \cite{ramakrishnan2019a}.
In addition, structural studies have shown
that phase transitions are accompanied by
charge ordering and spin dimerization \cite{radaellipg2002a}.
The occurrence of spin dimerization in a
three-dimensional compound, like CuIr$_2$S$_4$,
is unique and not understood at present.

\begin{figure}
\centering
\includegraphics[width=0.48\textwidth]{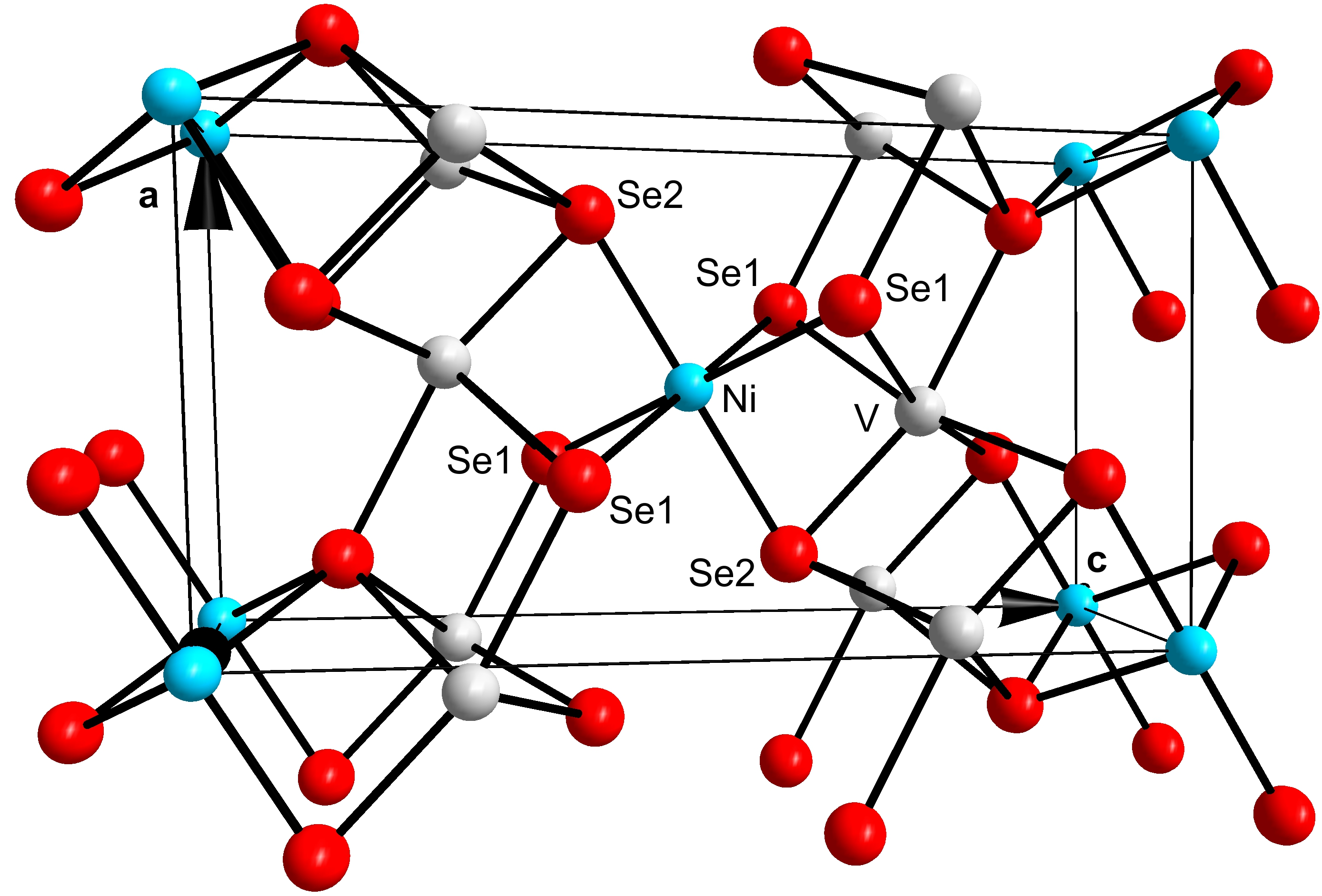}
\caption{Crystal structure of Ni$_{0.85}$V$_{2}$Se$_{4}$
in a perspective view along ${\mathbf{b}}$.
The shortest distances between metal atoms form
an unbounded network on the $(1\,0\,1)$ plane;
$d$(Ni--V) = 2.988 (1) \AA{} and $d$(V--V) = 3.033 (1) \AA{}
at room temperature.
The next shortest metal distance is $b = 3.4327$ \AA{} along $[0, 1, 0]$.}
\label{f-niv2se4_basic_structure}
\end{figure}

Apart from these cubic spinel compounds,
there exists more than 50 chalcogenides
with the same stoichiometry AT$_2$X$_4$
(A = Y, Cr, Fe, Co, Ni; T = Ti, Y, Cr; X = S, Se, Te),
but crystallizing in the monoclinic
defect NiAs structure type
(\textbf{Figure \ref{f-niv2se4_basic_structure}}) \cite{wolda1993a},
where very little investigations have been made.
When we consider the band structures of
compounds with the defect NiAs structure,
we find that only octahedral site ions (both
Ni and other transition metal) are involved.
Unlike spinels, all metal ions are on
octahedral sites and, therefore, the
$t_{2g}$ levels have the lowest energy.
The metallic behavior for the AT$_2$X$_4$
compounds can be ascribed to partially
filled bands formed as a result of T--T
overlap.
The NiAs structure is also known as the
Cr$_3$S$_4$ structure type (space group
$I2/m$) with two nonequivalent
crystallographic sites for Cr.
Chromium atoms occupy Wyckoff positions
2a and 4i, with two and four atoms per
unit cell, respectively.
These sites can be occupied by different
elements \cite{bouchardrj1969a, jellinekf1957a, powellav1997a, powellav1999a, vaqueirop2001a,plovnickrh1968a,morrisbl1970a, morrisbl1969a, vaqueirop2000a}.

A study on polycrystalline NiV$_2$S$_4$
has indicated a CDW transition at 165 K
\cite{kuoyk2007a}, while another study on
NiV$_2$Se$_4$ reported a small anomaly in
the resistivity around 160 K \cite{bouchardrj1966a}.
Using the vapor transport method, we have
obtained a Ni-deficient single crystal of
composition Ni$_{0.85}$V$_{2}$Se$_{4}$,
which has the same crystal structure as
NiV$_2$Se$_4$.
Band structure calculations reveal that it
is a $Z_2$  topological metal.
Our experimental studies on the physical
properties of Ni$_{0.85}$V$_{2}$Se$_{4}$
single crystal unfold an unusual electronic
ground state of this material with exotic
properties.

\section{\label{sec:niv2se4_discussion}%
Results and discussion}

\subsection{\label{sec:niv2se4_discussion_xray}%
Evidence of vacancies and absence of lattice distortion.}

Single-crystal x-ray diffraction (SXRD)
data were processed by Eval15 \cite{schreursamm2010a}
and SADABS \cite{sheldrick2008}.
All diffraction maxima could be indexed by a
single unit cell with lattice parameters
$a = 6.1453(3)$ \AA{}, $b = 3.4327(2)$ \AA{},
$c = 11.5901(4)$ \AA{} and $\beta = 91.619(3)$ deg,
similar to the published unit
cell of NiV$_2$Se$_4$ \cite{bouchardrj1966b}.
Crystal structure refinements were done with
Jana2006 \cite{petricekv2014a} with the
monoclinic space group $I2/m$.
\textbf{Table \ref{tab:niv2se4_compare}} shows the
structural analysis for three different
models on basis of the occupancy of the
Ni site, in order to determine whether
there are vacancies on the Ni site or
there exists Ni/V disorder.
\begin{table}
\centering
\caption{\label{tab:niv2se4_compare}%
Summary of results of refinements against
SXRD data at 298K for three structure models,
differing in the treatment of the occupancy
of the Ni site.}
\begin{tabular}{c c c c}
\hline
Model               &I      & II       & III \\
\hline
Composition         & NiV$_{2}$Se$_4$  & Ni$_{0.4178(143)}$V$_{2.5822}$Se$_4$   &  Ni$_{0.8524(35)}$V$_{2}$Se$_{4}$ \\
Occ[Ni]            &  1   & 0.4178(143) &  0.8524(35) \\
Occ[V]             &  1   & 1           &  1   \\
Occ[V$_{Ni}$]      &  -   & 0.5822      &  -   \\
Unique reflections (obs/all) & 578/685 & 578/685  & 578/685 \\
No. of parameters  & 24         &   25      & 25 \\
$R_{F}$(obs)       & 0.0317     & 0.0213    & 0.0209 \\
$wR_{F}$(all)      & 0.0454     & 0.0253    & 0.0255 \\
GoF (obs/all)      & 2.65/2.45  & 1.46/1.37 &  1.43/1.34 \\
$\Delta\rho_{min}$, $\Delta\rho_{max}$(e \AA$^{-3})$ &
-2.87, 1.49  & -1.07, 1.30   & -1.08, 1.28 \\
\hline
\end{tabular}
\end{table}

Initially, model I was tested where
the Ni site is completely filled.
The composition is NiV$_{2}$Se$_4$ and
it leads to a good fit of the
diffraction data with $R = 0.0317$.
In model II, a Vanadium atom (V$_{Ni}$)
was introduced on the Ni site, such that
the sum of the occupancies remains 1,
Occ[Ni] + Occ[V$_{Ni}$] = 1.
Refinement of Occ[V$_{Ni}$] then leads to
a significant improvement of the fit, with
$R = 0.0213$.
However, the chemical composition derived
from model II (Table \ref{tab:niv2se4_compare})
is not in agreement with the composition \\
Ni$_{0.925(9)}$V$_{2.054(7)}$Se$_{4}$ found
by an electron probe micro-analyzer (EPMA).
Such a large difference in the measured
compositions between SXRD and EPMA led us
to believe that the partially filled site
of Ni is not occupied by V.
Therefore, model II is not suitable.
Model III involves vacancies at the Ni site.
Refinement of the occupancy of Ni results in
an even better fit to the data, with $R = 0.0209$.
Moreover the chemical composition of model III
is in agreement with that obtained by EPMA,
with a difference of about 7\% deemed
acceptable.
One can also notice the standard uncertainty (su.)
in the refined occupancy of Ni in model III
is four times smaller than in model II.
Our findings from SXRD and EPMA suggest that
there is 15\% vacancies on the Ni site.
\textbf{Table \ref{tab:niv2se4_atom}} provides the
atomic coordinates and atomic displacement
parameters (ADPs) of model III.
\begin{table}
\centering
\caption{\label{tab:niv2se4_atom} %
Structural parameters for the monoclinic
structure of crystal A of Ni$_{0.85}$V$_{2}$Se$_4$  at 298 K.
Given are the fractional coordinates $x$, $y$, $z$ of the atoms,
their anisotropic displacement parameters (ADPs)
$U_{ij}$ $(i, j = 1, 2, 3)$
and the equivalent isotropic displacement parameter $U^{eq}_{iso}$. All atomic sites, except Ni, are fully occupied.}
\begin{tabular}{cccccccccccc}
\hline
Atom & Occ  & $x$ & $y$ & $z$ & $U_{11}$ & $U_{22}$ & $U_{33}$ & $U_{12}$ & $U_{13}$ & $U_{23}$ & $U^{eq}_{iso}$ \\
\hline
Ni &0.8524(35) &0 & 0 & 0 &  0.0158(3) &   0.0019(4) &  0.0017(4) &  0 & 0.0011(3) & 0 &   0.0175(2) \\
V &1 &-0.0451(1) &0 &0.2556(1) &  0.0133(2) &   0.0021(3) &  0.0153(3) &   0 & -0.0006(2) & 0 &  0.0166(2) \\
Se1 &1 &0.3373(1) &0 &0.3641(1) &   0.0119(2) & 0.0146(1) &  0.0176(2) &0 &-0.0003(1) &0 &   0.0015(1) \\
Se2 &1 &0.3381(1) &0 &0.8924(1) &    0.0138(1) &  0.0153(2) &  0.0163(2) &0 &-0.0004(1) &0 &  0.0152(1) \\
\hline
\end{tabular}
\end{table}

From the measured diffraction data at 298 K
and at 100 K, reciprocal layers were reconstructed
for $0kl$, $hk0$ and $h0l$ (Figure S1, S2 and S3
in the Supplemental Material \cite{niv2se4suppmat2022a}).
They demonstrate that,
upon cooling the crystal down to 100 K, there are no
superlattice reflections observed and there is
no change to the lattice, indicating the absence
of a CDW phase transition.
\textbf{Table \ref{tab:niv2se4 crystalinfo}} shows the
crystallographic information of the crystal at
298 and 100 K, respectively, as based on model III.
\begin{table}
\caption{\label{tab:niv2se4 crystalinfo}%
Crystallographic data of crystal A of
Ni$_{0.85}$V$_{2}$Se$_4$ at 298 K and 100 K.
Notice that space group $I2/m$ is a non-standard
setting of space group No. 12 with standard
setting $C2/m$ \cite{itvola}.}
\centering
\begin{tabular}{c c c}
\hline
Temperature (K) & 298 & 100  \\
\hline
Crystal system & Monoclinic& Monoclinic \\
Space group & $I2/m$  &$I2/m$ \\
Space group no. & 12 & 12   \\
$a$ (\AA{}) &6.1453(3) &6.1314(2)  \\
$b$ (\AA{}) &3.4327(2)    &3.4174(3)     \\
$c$ (\AA{}) &11.5901(4)    &11.5462(3)  \\
$\beta$ (deg) & 91.619(3) & 91.741(3) \\
Volume (\AA{}$^3$) & 244.45(3) & 241.83(3)  \\
$Z$ & 2 & 2 \\
Wavelength (\AA{}) & Ag-K$\alpha$ &Ag-K$\alpha$  \\
Detector distance (mm) &100 &100  \\
$2\theta$-offset (deg) &0 &0, 30  \\
Rotation per image (deg) & 1 & 1  \\
$(\sin(\theta)/\lambda)_{max}$ (\AA{}$^{-1}$) &0.683589& 1.009913 \\
Absorption, $\mu$ (mm$^{-1}$) & 19.271 & 19.469  \\
T$_{min}$, T$_{max}$ & 0.5796, 0.8622 & 0.6018 0.8627  \\
Criterion of observability & $I>3\sigma(I)$ & $I>3\sigma(I)$ \\
Number of  reflections \\
measured &  4410  & 8970  \\
unique (obs/all) & 451/470 &841/1024  \\
$R_{int}$  (obs/all) & 0.0312/0.0321 &0.0407/0.0415  \\
Composition     &  Ni$_{0.8524(35)}$V$_{2}$Se$_{4}$    &  Ni$_{0.8488(25)}$V$_{2}$Se$_{4}$\\
No. of parameters &25 &25 \\
Occ[Ni]    & 0.8524(35)       &  0.8488(25)   \\
$R_{F }$  (obs) & 0.0209  &0.0196 \\
$wR_{F }$ (all) &0.0255  &0.0246 \\
GoF (obs/all) &1.43/1.34 &1.32/1.21 \\
$\Delta\rho_{min}$, $\Delta\rho_{max}$(e \AA$^{-3}$) &
 -1.08, 1.28 &-1.27, 1.71  \\
\hline
\end{tabular}
\end{table}
For details regarding data processing
of the 100 K data refer Section S1 in the supplemental material \cite{niv2se4suppmat2022a}.
The absence of the CDW transition
is substantiated by our band structure calculations
described in Section \ref{sec:band structure}.

\subsection{\label{sec:band structure}%
Structural stability and electronic structure}

\begin{figure}
\centering
\includegraphics[width=0.95\textwidth]{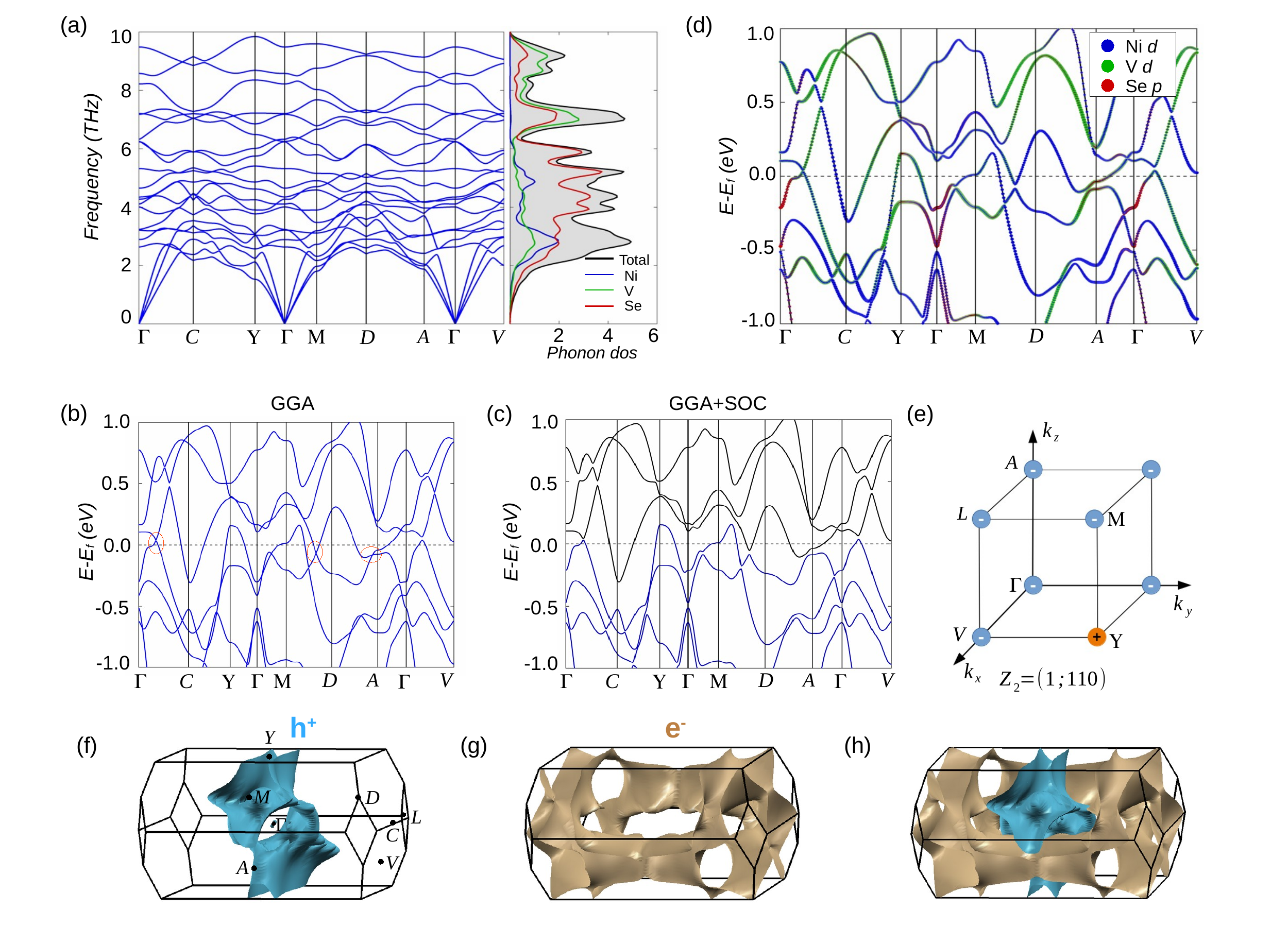}
\caption{(a) Calculated phonon dispersion and total and partial phonon density of states of bulk NiV$_2$Se$_4$.
The low-frequency phonons mainly arise from Ni and Se atoms whereas high-frequency phonons arise from V and Se atoms.
Bulk band structure (b) without and (c) with spin-orbit coupling (SOC).
The nodal band crossings along high-symmetry directions are highlighted with red circles in (b).
These crossing points are gapped in (c), leading to a local band gap between valence (blue lines) and conduction (black lines) bands.
(d) Orbital resolved band structure with SOC.
Blue, green, and red markers identify Ni $d$, V $d$, and Se $p$ states, respectively.
(e) Parity eigenvalues of the occupied bands at eight time-reversal invariant momentum points.
(f) and (g) Calculated individual Fermi pockets and (h) the Fermi surface of NiV$_2$Se$_4$.}
\label{f-niv2se4_band}
\end{figure}

We now discuss the stability and nontrivial state
of NiV$_2$Se$_4$ with regards to the calculated
band structure \textbf{(Figure \ref{f-niv2se4_band})}.
These results also apply to
Ni$_{0.85}$V$_{2}$Se$_{4}$, because the major
effect of the introduction of vacancies is
a shift of the Fermi level, which is presented
in Section S3 of the Supporting Information.
The results on Ni-deficient material show a 50 meV
downward shift of the
Fermi level as compared to pristine NiV$_2$Se$_4$.
The phonon dispersion curves and associated phonon
density of states (PDOS) are shown in
Figure \ref{f-niv2se4_band}(a).
The absence of imaginary phonon frequency in the
full BZ ensures that NiV$_2$Se$_4$ is dynamically
stable.
Moreover, the absence of a soft phonon mode
indicates the absence of a structural instability
or a periodic lattice distortion,
as it would be expected in CDW materials,
in agreement with the experiments.
The PDOS suggests that Ni and Se atoms contribute
to low-lying phonon modes, whereas the V and Se
atoms dominate the high-frequency phonon modes.

The bulk band structure along the high-symmetry
directions in the BZ of NiV$_2$Se$_4$
is shown in Figure \ref{f-niv2se4_band}(b)
without spin-orbit coupling (SOC).
It is metallic, where both the valence and
conduction bands cross the Fermi level.
Notably, the valence and conduction bands cross
along $\Gamma$--$C$, $M$--$D$, $D$--$A$
directions at generic $\mathbf{k}$ points
close to the Fermi level,
as indicated by red circles.
The crossing points are stable against band
hybridization to realize nodal line crossings.
The band structure in presence of SOC is
shown in Figure \ref{f-niv2se4_band}(c).
Although the system remains metallic, the band
crossing points are now gapped,
separating the valence and conduction bands
locally at each $\mathbf{k}$ point.
The orbital resolved band structure in
Figure \ref{f-niv2se4_band}(d) shows that
these band crossing points are composed of
Ni--$d$, V--$d$, and Se--$p$ orbitals.
The existence of local band gap allows
the computation of the $Z_2$ invariants $(\nu_0;\nu_1\nu_2\nu_3)$ in a manner
similar to $Z_2$ topological insulators.
Since NiV$_2$Se$_4$ crystal respects inversion
symmetry, we can calculate
the $Z_2$ invariants $(\nu_0;\nu_1\nu_2\nu_3)$
from the parity eigenvalues of the valence bands
at each time-reversal-invariant momentum (TRIM)
point \cite{fu2007topological}.
The product of the parity eigenvalues of the
occupied bands at the eight TRIM points is
shown in Figure \ref{f-niv2se4_band}(e).
The calculated $Z_2$ of $(1;110)$ indicates
that NiV$_2$Se$_4$ is $Z_2$ topological metal.
The topological phases are immune to the effects
of a Hubbard $U$ for the Ni and V atoms, as
discussed in Section S4 of the Supporting Information.

The nontrivial state is further demonstrated by
calculations of the surface electronic structure.
They reveal a single Dirac cone topological state
present within the energy gap of the bulk band
structure (see Section S3 of the Supplemental
Material for details \cite{niv2se4suppmat2022a}).
Figure \ref{f-niv2se4_band}(f)-(h) show the
calculated Fermi surface of NiV$_2$Se$_4$.
Owing to the multiband crossings at the Fermi
level, the Fermi surface consists of both
electron and hole bands which are shown
separately in Figure \ref{f-niv2se4_band}(f)
and \ref{f-niv2se4_band}(g), respectively.
The metallic behavior is consistent
with the experimental measurements.

\subsection{Electrical resistivity}

\begin{figure}
\centering
\includegraphics[width=0.48\textwidth]{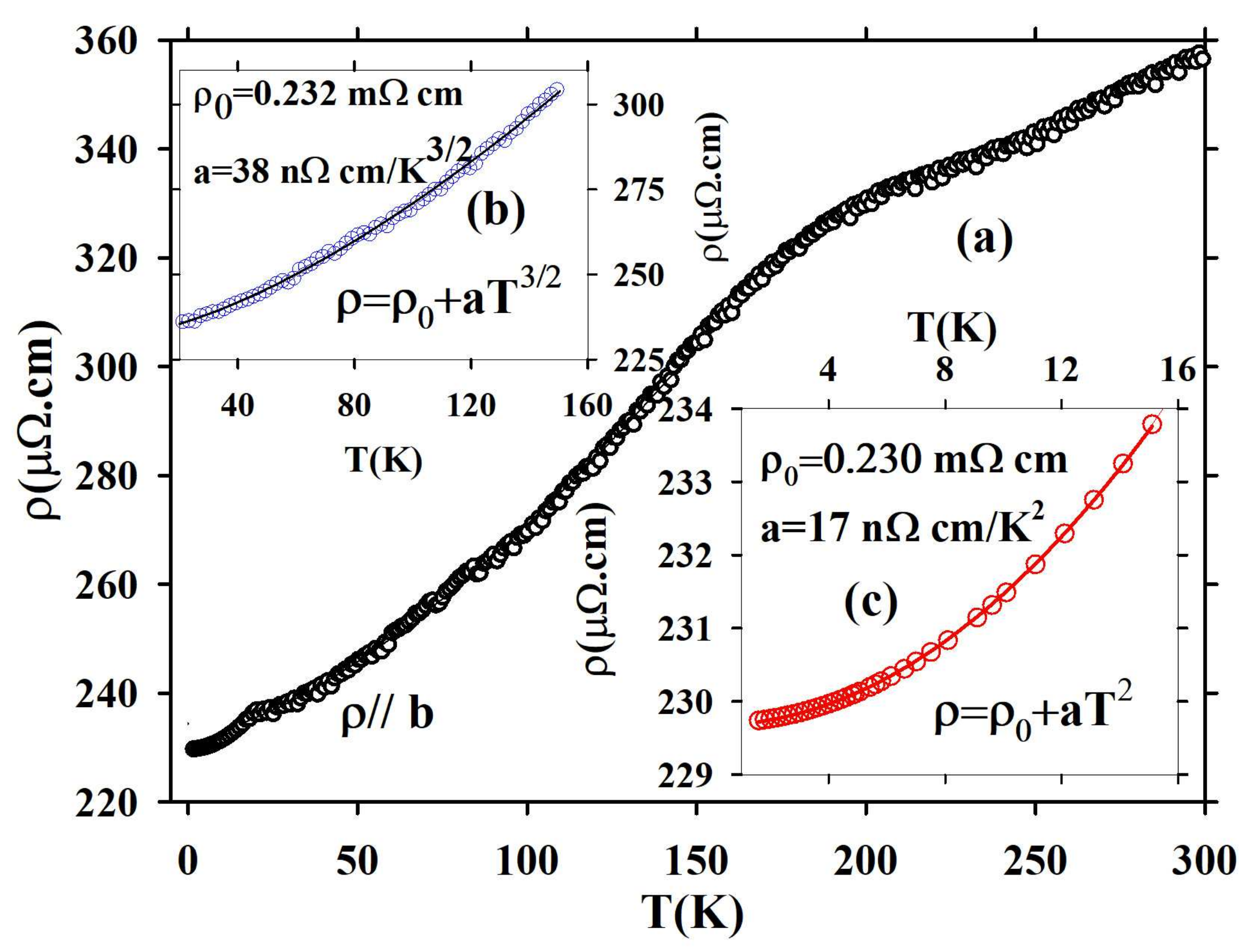}
\caption{Electrical resistivity of
Ni$_{0.85}$V$_{2}$Se$_{4}$.
Curve (a) $\rho(T)$ for the whole measured range
of temperatures 1.5--300 K.
Curve (b), upper inset, provides an expanded
view of 15--150 K;
the solid line is a fit to a $T^{3/2}$ dependence
(Equation \protect\ref{e-niv2se4_1rhot3half}), with
$\rho_{0,1} = 231.600\,(1)\; \mu\Omega\,$cm and
$a_1 = 0.038\,(1)\; \mu\Omega\,$cm/K$^{3/2}$.
Curve (c), lower inset, gives an expanded view
of 1.5--15 K; the solid line is a fit to a
$T^{2}$ dependence
(Equation \protect\ref{e-niv2se4_1rhot2}), with
$\rho_{0,2} = 229.982\,(1)\; \mu\Omega\,$cm and
$a_2 = 0.017\,(1)\; \mu\Omega\,$cm/K$^{2}$.}
\label{f-niv2se4_res}
\end{figure}

The electrical resistivity, $\rho(T)$, of Ni$_{0.85}$V$_{2}$Se$_{4}$
decreases upon cooling from room temperature down to 1.5 K
(\textbf{Figure \ref{f-niv2se4_res}}).
In particular, a broad hump is observed between
150--200 K,
which also agrees with Bouchard \textit{et al.}
(1966) \cite{bouchardrj1966a},
who first reported an anomaly at about 160 K in
$\rho(T)$ of a polycrystalline sample of NiV$_2$Se$_4$.
They suggested that the observed anomaly may be due
to either a crystallographic or a magnetic transition.
However, we find that the broad anomaly between
150--200 K cannot be ascribed to a structural
transition, since our diffraction studies did not
find any evidence for it
(Section \ref{sec:niv2se4_discussion_xray}). 
Explanations for these features could be
the development of spin fluctuations or 
CDW fluctuations as opposed to long-range 
magnetic order or CDW ordering.
The latter involve structural distortions, like
in CuV$_2$S$_4$ \cite{ramakrishnan2019a},
BaFe$_2$Al$_9$ \cite{meierwr2021a},
and $R_2$Ir$_3$Si$_5$ ($R$ = Lu, Er) \cite{ramakrishnan2021a,ramakrishnan2020a}.

The metallic behavior of Ni$_{0.85}$V$_{2}$Se$_{4}$
is consistent with the
properties of polycrystalline materials of
the defect-NiAs structure \cite{bouchardrj1969a,jellinekf1957a,powellav1997a,
powellav1999a,vaqueirop2001a,plovnickrh1968a,morrisbl1970a,morrisbl1969a,vaqueirop2000a,kuoyk2007a}.
Below the broad hump, $\rho(T)$ exhibits an
unusual $T^{3/2}$ dependence for 15--150 K.
Such a behavior is generally observed in spin
glasses and amorphous ferromagnets, where a
$T^{3/2}$ dependence in $\rho(T)$ is the result
of a diffusive motion of the charge carriers
\cite{hertzja1976a,millisaj1993a}.
Most theories suggest an explanation of the
anomalous resistivity by an inhomogeneous
magnetic state, through comparison with the
resistivity of spin glasses \cite{mydoshja1993a}.
Since Ni$_{0.85}$V$_{2}$Se$_{4}$
is a single crystal, the diffusive carrier motion
has to be intrinsic somewhat similar to the
behavior observed in MnSi under pressure
\cite{pfleidererc1997a}.

It is well established that the electronic
properties of metals on the border of magnetic
phase transitions at low temperatures are often
found to exhibit temperature dependencies that
differ from the predictions of Fermi liquid theory.
Early attempts to explore such non-Fermi-liquid
behavior have been based on mean-field treatments
of the effects of enhanced magnetic fluctuations,
as in the self-consistent renormalization (SCR)
model \cite{niklowitz2005a}.
The SCR model seems to explain the temperature
dependence of the resistivity of Ni$_3$Al at
high pressures, where its ferromagnetism is
suppressed, leading to a negligible magnetic
correlation vector $\kappa(T)$.
In the idealized limit $\kappa\rightarrow 0$
and $T\rightarrow 0$, the SCR model predicts
that in 3D, the quasiparticle scattering rate
$(\tau_{qp})^{-1}$ varies linearly with the
quasiparticle excitation energy, rather than
quadratically as in the standard Fermi-liquid picture.
This form of $(\tau_{qp})^{-1}$ is similar to
that of the marginal Fermi-liquid model
\cite{varmacm1989a}, which is normally
associated with a linear temperature dependence
of the resistivity.
However, at the border of ferromagnetism, the
relevant fluctuations responsible for
quasiparticle scattering are of long wavelength
and, thus, are ineffective in reducing the current.
This leads to a transport relaxation rate
$(\tau_{tr})^{-1}$ that differs from
$(\tau_{qp})^{-1}$ and is not linear in $T$.
Instead, it varies as $T^{5/3}$, which was
observed in high pressure studies on Ni$_3$Al.

For a metal on the border of metallic
antiferromagnetism in three dimensions,
the SCR model predicts $\rho(T)$ to vary as
$T^{3/2}$ in the idealized limit
$\kappa \rightarrow 0$, $T \rightarrow 0$,
where $\kappa(T)$ now stands for the correlation
wave vector for the staggered magnetization.
This simple model for the scattering from
antiferromagnetic fluctuations assumes that
the scattering rate can be averaged over the
Fermi surface.
Within the SCR model, this procedure would seem
to require the presence of a sufficient level
of defects, the latter which might be provided
by the chemical disorder due to vacancies at
the A-site in Ni$_{0.85}$V$_{2}$Se$_{4}$.
This site vacancy is supposed to inhibit the
short circuiting caused by the carriers on the
cold spots of the Fermi surface, \textit{i.e.},
regions far from the hot spots connected by the
antiferromagnetic ordering wave vector and, thus,
strongly affected by spin-fluctuation scattering.
Such a behavior has been observed in NiSSe \cite{miyasakas2000a}.

In the case of single crystalline
Ni$_{0.85}$V$_{2}$Se$_{4}$,
the temperature dependence of the electrical
resistivity between 15 and 150 K could be
fitted by
(upper inset (b) of Figure \ref{f-niv2se4_res})
\begin{equation}
\rho(T) = \rho_{0,1} + a_1 \,  T^{3/2} \, ,
\label{e-niv2se4_1rhot3half}
\end{equation}
where $\rho_{0,1}$ is the residual resistivity
and the second term is the contribution from
scattering proposed by the SCR model.
It is of interest to see that below 15 K,
$\rho(T)$ can be described by
(lower inset (c) of Figure \ref{f-niv2se4_res})
\begin{equation}
\rho(T) = \rho_{0,2} + a_2 \, T^{2} \, ,
\label{e-niv2se4_1rhot2}
\end{equation}
which suggests that the anomalous temperature
dependence ceases to exist below 15 K.
The second term in Equation \ref{e-niv2se4_1rhot2}
arises from the scattering of electrons due to
localized spin fluctuations \cite{kaiser1992}.

\subsection{Magnetic susceptiblity}

\begin{figure}
\centering
\includegraphics[width=0.48\textwidth]{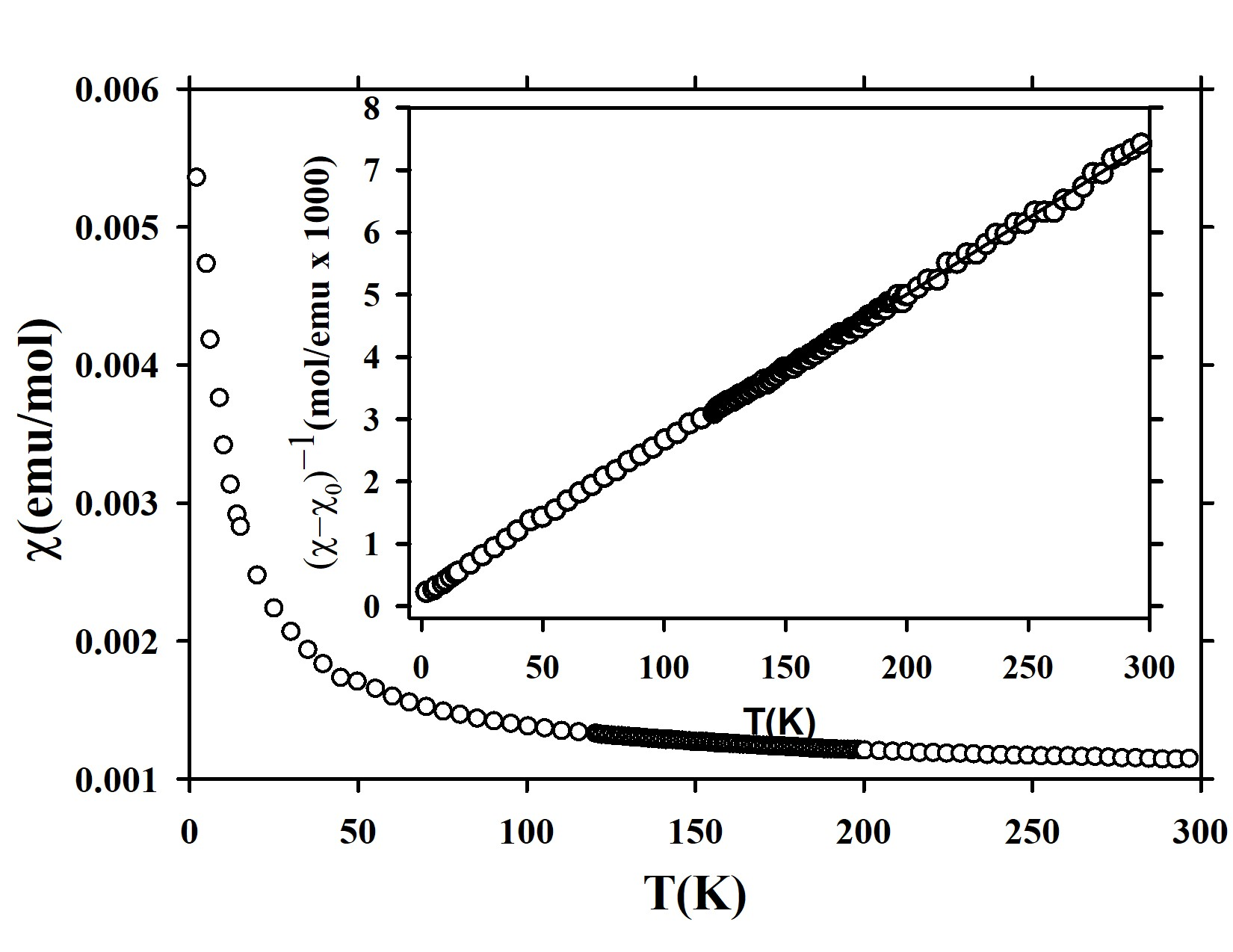}
\caption{Temperature dependence of the magnetic
susceptibility of Ni$_{0.85}$V$_{2}$Se$_{4}$.
Data measured in a field of 5\ T under zero-field
cooled (ZFC) conditions.
A measurement under field cooled (FC) conditions
produced data coinciding with those shown for ZFC
conditions.
The inset provides the inverse susceptibility;
the solid line is a fit to the modified Curie-Weiss
law (Equation \protect\ref{e-niv2se4_curieweiss}),
with $\chi_0 = 1.01\,(1)\times 10^{-3}\,$emu/mol,
$C = 0.040\,(1)\,$emu/mol$\,$K,
$\mu_{eff} = 0.57\,(1)\,\mu_{B}$
and $\theta = 4.8\,(1)\,$K.
}
\label{f-niv2se4_ZFC}
\end{figure}

The temperature dependence of the magnetic
susceptibility clearly shows paramagnetic behavior
(\textbf{Figure \ref{f-niv2se4_ZFC}}).
However, it does deviate from a simple
Curie-Weiss behaviour, as discussed below.
Magnetization measurements performed under
field cooled (FC; not shown) and zero-field
cooled (ZFC; Figure \ref{f-niv2se4_ZFC})
conditions produced identical results,
implying that spin glass phenomena can be ruled
out in the case of Ni$_{0.85}$V$_{2}$Se$_{4}$.
The high temperature (150--300 K) data can be
described by a modified Curie-Weiss law,
\begin{equation}
\chi = \chi_0 + C/(T + \theta) \, .
\label{e-niv2se4_curieweiss}
\end{equation}
where $\chi_0$ includes the diamagnetic and temperature-independent paramagnetic contributions.
From the value of the Curie constant $C$,
the effective magnetic moment is estimated as
$\mu_{eff} = 0.57\,(1)\; \mu_B$ per formula unit.
Usually, the Ni ion does not have a magnetic moment
in metallic compounds of the type AT$_2$X$_4$,
since all valence electrons of Ni are completely
delocalized in fully ordered crystal structures.
Apparently, this scenario breaks down in Ni$_{0.85}$V$_{2}$Se$_{4}$, which again might be
related to the vacancies on the Ni site.
The Weiss temperature is found to be
$\theta = 4.8\,(1)$ K, indicative of weak
antiferromagnetic exchange interactions.
The inverse susceptibility (Figure \ref{f-niv2se4_ZFC})
shows small deviations from linear behavior
below 150 K.

\begin{figure}
\centering
\includegraphics[width=0.48\textwidth]{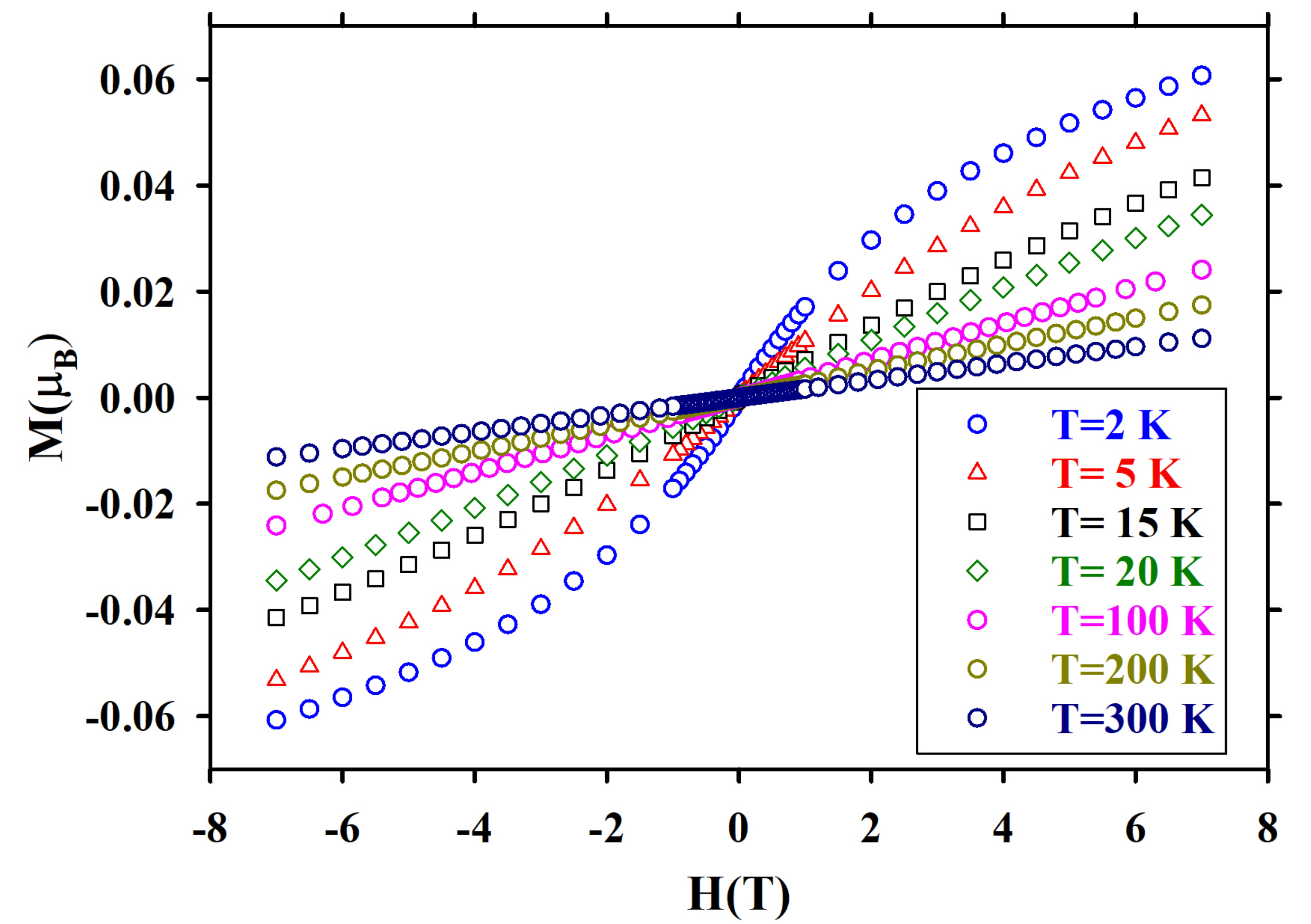}
\caption{Magnetization (M) in dependence on the applied
magnetic field (H) at selected temperatures.}
\label{f-niv2se4_MH}
\end{figure}

Isothermal magnetization (M--H) curves have been
measured at selected temperatures within the
range 2--300 K up to magnetic fields of
$\pm 7$\ T (\textbf{Figure \ref{f-niv2se4_MH}}).
Hysteresis is absent at all temperatures,
demonstrating that Ni$_{0.85}$V$_{2}$Se$_{4}$
does not undergo a magnetic phase transition,
but remains in the paramagnetic state down to 2 K.
At the lowest measured temperatures, the M--H
dependence is S-shaped instead of linear at
higher temperatures.
This might be related to the fact that lower
temperatures allow to reach higher values of H/T,
thus bringing the M--H relation outside the
linear regime.
However, attempts failed to bring all data onto
a universal H/T dependence.
This seems to imply that the crystal does
not behave like an usual paramagnet below 15 K.

\subsection{Specific heat}

\begin{figure}
\centering
\includegraphics[width=0.48\textwidth]{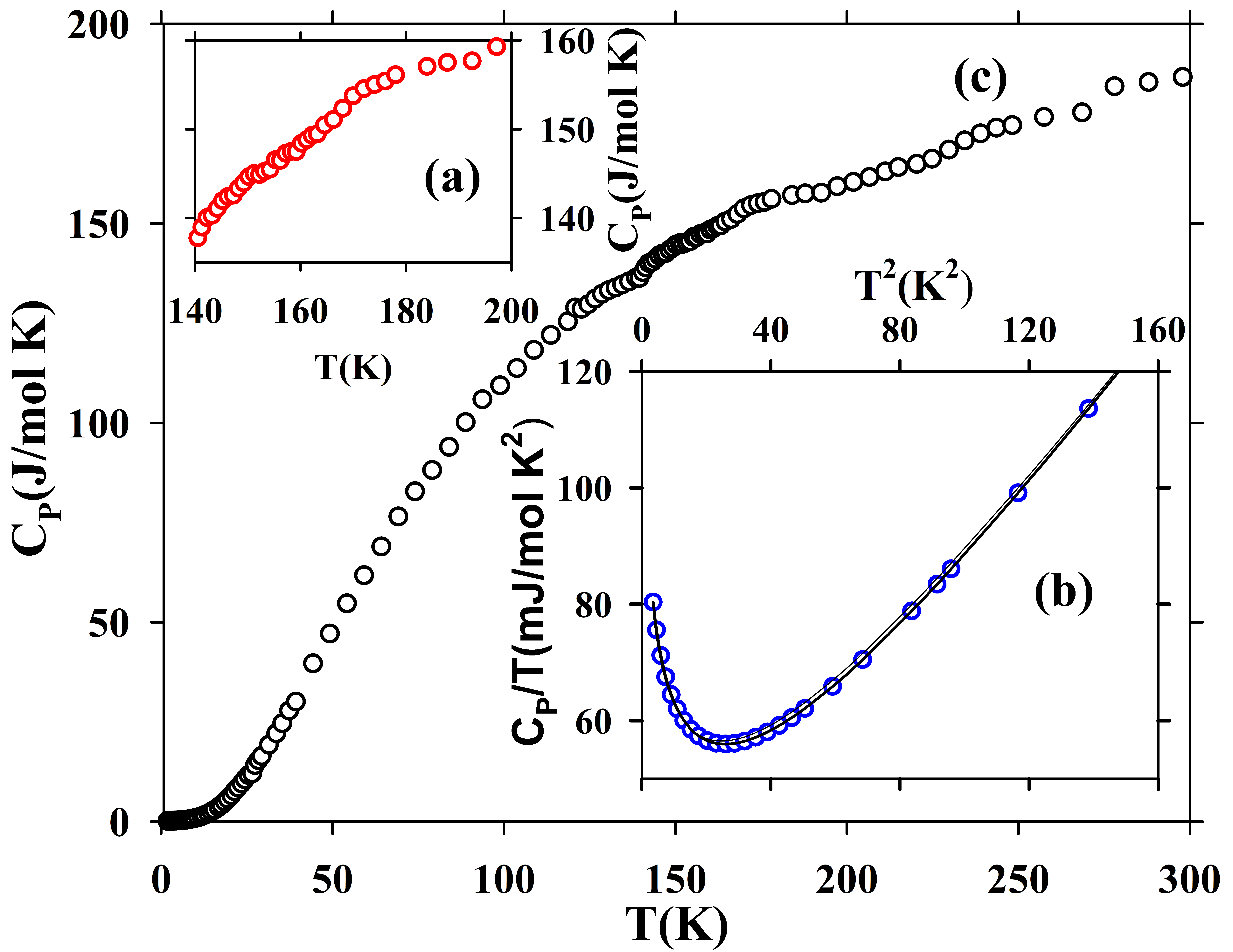}
\caption{Temperature dependence of the specific
heat $C_p$.
(c) displays $C_p(T)$ for temperatures 2--300 K.
The upper inset (a) provides an expanded view of
the region of the anomaly around 160 K.
The lower inset (b) displays $C_p/T$ \textit{vs}
$T^2$ at low temperatures 2--12 K.
Contributions from strong spin fluctuations are
demonstrated by a fit (solid line) of
Equation \protect\ref{e-niv2se4_1hc} to the data,
resulting in $\gamma = 104.0\,(1)$ mJ\,mol$^{-1}$\,K$^{-2}$,
$\beta = 0.83\,(1)$ mJ\,mol$^{-1}$\,K$^{-4}$,
$\theta_D = 252.9\,(1)$ K,
$a = 20.1\,(1)$ mJ\,mol$^{-1}$\,K$^{-2}$ and
$T^{*}=0.86\,(1)$ K.}
\label{f-niv2se4_hc}
\end{figure}

The specific heat, $C_p(T)$, decreases on cooling
(\textbf{Figure \ref{f-niv2se4_hc}}).
The absence of sharp features and the presence
of a broad, weak anomaly at $T$ = 150--170 K rule
out a possible structural phase transition,
which is in agreement with our single-crystal
x-ray diffraction data.
However, the presence of this anomaly in the
specific heat data is in agreement with the
anomalies observed in resistivity data.
The origin of this anomaly could be attributed to
spin or CDW fluctuations.
It is important to specify here, that the
anomalous temperature dependence of the
resistivity ($T^{3/2}$) starts below this transition.
The low temperature data can be fitted with
the model \cite{lonchakov2014a},
\begin{equation}
C_p/T = \gamma + \beta \, T^{2} + a \,  \ln(T^{*}/T) \, .
\label{e-niv2se4_1hc}
\end{equation}
The first term is the contribution due to
conduction electrons, the second term is the
contribution from the lattice, and the third
term is the contribution from spin
fluctuations due to disorder.
This model was successfully employed to explain
a similar anomaly in the specific heat of
Fe$_{1-x}$V$_{2+x}$Al
in another context \cite{lonchakov2014a}.
The fit yields $\gamma = 104.0\,(1)$ mJ/mol$\,$K$^2$
and the Debye temperature $\theta_D = 252.9\,(1)$ K.
The existence of magnetic fluctuations below
15 K has been corroborated by the isothermal
magnetization data, as explained earlier.
The presence of such magnetic fluctuations seems
to change the anomalous
temperature dependence of resistivity behavior below 15 K.

\section{Conclusions}

The perfectly ordered crystal structure of
NiV$_2$Se$_4$ does not have a magnetic moment on Ni.
However, our attempted growth of single crystals of NiV$_2$Se$_4$ has resulted in Ni-deficient
crystals, where the crystallographic Ni site has
15\% vacancies, but full chemical order is preserved
at the V and the two Se sites of the
AT$_2$Se$_4$ monoclinic structure type.
This observation is in agreement with reports in
the literature of crystallization attempts,
where part or all of the transition metal A can
be occupied by defects or V atoms
\cite{bouchardrj1966a,kallela1984a}.

Single crystalline Ni$_{0.85}$V$_{2}$Se$_{4}$
is found to be a $Z_2$ topological metal,
that exhibits a cross-over in its physical
properties at temperatures around 150 K.
All properties, including low-temperature
x-ray diffraction data, are in agreement
with the absence of phase transitions down
to $T = 1.5$ K in the absence of a magnetic field.
The temperature dependence of the electrical
resistivity of Ni$_{0.85}$V$_{2}$Se$_{4}$ is
anomalous in the
temperature range from 15--150 K,
while the presence of magnetic correlations below 15 K probably leads to T$^2$ dependence of resistivity.
It is interesting to compare the present results
to the well established $Z_2$ topological metal
$R$V$_6$Sn$_6$ ($R$ = Gd and Y) \cite{ganesha2021a},
containing a kagome network of V ions coordinated
by Sn and separated by triangular lattice planes
of rare-earth ions.
The Gd spins contribute to the observed magnetism
of GdV$_6$Sn$_6$ while YV$_6$Sn$_6$ is in a
paramagnetic state.
Unlike the case of YV$_6$Sn$_6$, minor site
disorder of Ni in Ni$_{0.85}$V$_{2}$Se$_{4}$
is responsible for its anomalous low-temperature
properties.
The broad hump in the resistivity may be related
to the anomaly in the specific heat in the
temperature range 150--160 K.
However, the exact nature of the crossover is not
understood at present.
These features combine with the gradual build-up 
of antiferromagnetic fluctuations
upon cooling.
Isothermal magnetization curves confirm the
absence of bulk magnetic order down to 2 K,
but attempts have failed to bring all data
onto a universal H/T dependence.
This seems to imply that the crystal does not
behave like a usual paramagnet below 15 K,
and it suggests a build-up of antiferromagnetic
correlations.
We believe that Ni$_{0.85}$V$_{2}$Se$_{4}$ is a $Z_2$
is a topological metal that displays unusual
electronic properties due to significant electron
correlations.
However, to understand this system,
one needs more investigations,
notably on the nature of the anomalous behavior
around 160 K and
Fermi surface studies using angle-resolved
photoemission spectroscopy.


\section{\label{sec:niv2se4_experimental}%
Experimental and computational section}

\threesubsection{Crystal growth and characterization by EPMA}\\
Attempts to synthesize single-crystalline NiV$_{2}$Se$_4$ resulted in the formation of Ni$_{0.85}$V$_{2}$Se$_{4}$
prepared by solid-state reaction of the elements in evacuated quartz-glass
tubes at a temperature of $T = 1153$ K, employing stoichiometric amounts of the elements obtained
from Alfa Aesar: nickel (99.996\% purity), vanadium (99.5\%),
and selenium (99.999\%).
Single crystals have been grown by
vapor transport in evacuated quartz-glass tubes, employing
iodine as transport agent and a temperature gradient
of $T = 1033\,/\,973$ K over 200 mm.
The chemical composition has been determined as
Ni$_{0.925(9)}$V$_{2.054(7)}$Se$_{4}$ by an electron
probe micro-analyzer (EPMA). These values are the
average over 57 points measured on a a flat,
as-grown surface sliced off from the large single crystal
used for the physical-properties measurements. The finding
agrees with the results of x-ray diffraction
(see Section \ref{sec:niv2se4_discussion_xray}).

\threesubsection{Temperature dependent Single-crystal
x-ray diffraction (SXRD)}\\
SXRD has been measured on a mar345dtb
image plate diffractometer (marXperts, Germany)
with Ag-K$\alpha$ radiation from a rotating
anode generator.
Initially, several crystals from
the batch were tested, all showing that they were Ni deficient
similar to what has been found with EPMA.
For the final reported measurement a crystal
of dimensions $0.1\times 0.1\times 0.1$ mm$^{3}$
was crushed from the large crystal which was used for
physical properties measurements.

A complete set of diffraction images has been
measured at temperatures 298 K and 100 K,
employing an exposure time such that the strongest
Bragg's reflection was close to the saturation of the detector.
A second
run was measured with an offset of 30 deg in $2\theta$ for obtaining
data at high resolution, and employing an eight times longer exposure
time, resulting in overexposed strong reflections but allowing for
detection of weak scattering effects. Superlattice reflections could
not be detected, in agreement with the absence of a CDW at 100 K.
Details are given in Section S1 and S2 of
the Supplemental Material \cite{niv2se4suppmat2022a}.

\threesubsection{Physical property measurements}\\
The electrical resistivity $\rho(T)$ has been measured on cooling
from 300 down to 1.5\ K in a standard four-probe configuration,
employing a cryostat and a LR-700 (Linear Research, USA) bridge
with 5 mA current of small AC frequency of 16 Hz. Electrical
contacts have been made using silver paste and gold wires ($40\;\mu$m diameter).
The magnetic susceptibility $\chi(T)$ has been measured for
temperatures 2--300\ K, using a commercial SQUID magnetometer
(MPMS5 by Quantum Design, USA). Measurements have been repeated
in both zero-field-cooled (ZFC) and field-cooled (FC) conditions
for magnetic fields of different strength. The isothermal
magnetization has been measured for magnetic fields
from ${-}7$ to +7 Tesla at a few selected temperatures.
The heat capacity $C_{p}(T)$ has been measured from
2 to 300 K by the thermal relaxation method using a
Physical property measuring system (PPMS, Quantum Design, USA).

\threesubsection{Density functional theory calculations}\\
Electronic structure calculations were performed using the Vienna ab
initio simulation package (VASP)~\cite{kresse1996efficient}
with the projector augmented (PAW)~\cite{blochl1994projector}
wave method. Generalized gradient approximation of Perdew,
Burke and Ernzerhof (PBE)~\cite{perdew1996generalized} was used
to include the exchange-correlation effects. An energy  cutoff
of 380 eV was used for the plane wave basis and a
$10\times 10 \times 10$ $\Gamma$ centered k-mesh was employed for
the Brillouin zone (BZ) sampling. Experimental lattice parameters
with fully optimized ionic positions were considered in our computations.
We optimized atomic positions until the residual
forces on each atom were less than 0.0001 eV/{\AA}.
The material-specific tight-binding Hamiltonian ~\cite{marzari1997maximally, arash2008wannier90}
was generated to compute the topological properties using the
WannierTools ~\cite{quansheng2018wanniertools} package.
The Xcrysden program was used to visualize the Fermi surface~\cite{anton1999xcrysden}.
The phonon spectrum was obtained using the frozen phonon method as implemented in the
Phonopy~\cite{togo2015first} code with a $2\times 2 \times 2$ supercell.

\medskip
\textbf{Supporting Information} \par 
Supporting Information is available from the
Wiley Online Library or from the author.

\medskip
\textbf{Acknowledgements} \par 
Single crystals have been grown by
Kerstin K\"{u}spert at the Laboratory of
Crystallography in Bayreuth.
We thank Detlef Krau\ss{}e of the Bavarian Geoinstitute
(BGI) in Bayreuth for performing the electron microprobe experiments.
This research has been funded by the Deutsche
Forschungsgemeinschaft
(DFG, German Research Foundation)–265092781.
SM thanks the Tata Institute of Fundamental Research Mumbai
for providing a postdoctoral fellowship and VGST CESEM GRD-852.
The work at TIFR is supported
by the Department of Atomic Energy of the Government
of India under Project No. 12-R\&D-TFR-5.10-0100.

\medskip
\textbf{Conflict of interest} \par 
The authors declare no conflict of interest

\medskip
\bibliographystyle{MSP}
\bibliography{Ref1.bib}


\vspace{40mm}
\begin{figure}[!h]
\textbf{Table of Contents}\\
\medskip
\includegraphics[height=54.892mm]{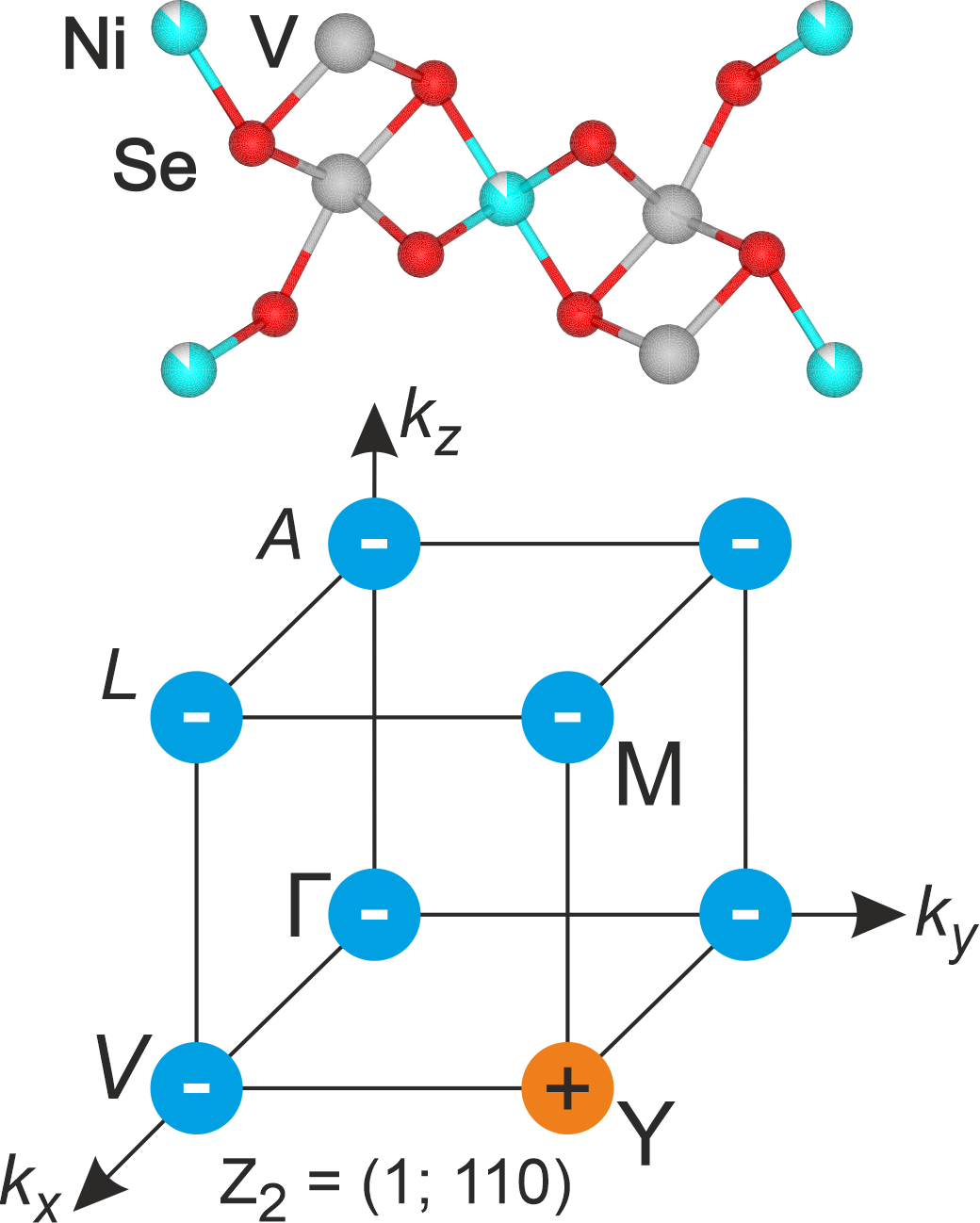}
\medskip
\caption*{Growth of single crystals of nominal
NiV$_2$Se$_4$ has produced nickel-deficient
Ni$_{0.85}$V$_2$Se$_4$.
Vacancies at the nickel site explain the absence
of a charge-density-wave (CDW) transition, while
the temperature
dependencies of physical properties exhibit
broad anomalies, indicative of
spin or CDW fluctuations. 
Density-Functional-Theory (DFT) calculations
reveal that Ni$_{0.85}$V$_2$Se$_4$ is a $Z_2$
topological metal. }
\end{figure}

\end{document}